\documentclass[showkeys,showpacs]{revtex4}
\usepackage{graphicx}
\usepackage{amsmath}
\usepackage{amssymb}
\usepackage{color}

\begin{document}
\title{Nonperturbative quantization and turbulence: the comparison}

\author{Vladimir Dzhunushaliev}
\email{v.dzhunushaliev@gmail.com}
\affiliation{
Dept. Theor. and Nucl. Phys., Kazakh National University, Almaty, 010008, Kazakhstan}

\begin{abstract}
The comparison of calculations methods in nonperturbative quantum fields theory and turbulence theory is made. The main result is that in both cases there is an infinite equations set. In the first case it is the equations set for Green's functions and in the second case it is the equations set for cumulants. It allows us to use similar mathematical methods to solve problems in nonperturbative quantum field theory and turbulence modeling. A closure problem of truncation of the infinite equations set is discussed.
\end{abstract}


\maketitle

\newpage

\section{Introduction}

Unfortunately up to now the quantization of strongly interacting fields and the turbulence theory remain unsolved problems. For example, the problem of quantum chromodynamics and gravity is that Feynman diagram technique can not be applied for quantizing of SU(3) non-Abelian field and gravity. Physically it means that weakly interacting quantum particles interact with each other at the vertices but it is not the case for strongly interacting fields. It do not allows us to use Feynman diagram technique which works with quantum particles interacting at the vertices. The problem with the turbulence is that we have not yet proven that in three dimensions solutions always exist, or that they do not contain any singularity (so called the Navier - Stokes existence and smoothness problems).

Probably at the first time the nonperturbative quantization was applied by Heisenberg to quantize a nonlinear spinor field \cite{heisenberg}. The procedure of nonperturbative quantization procedure is much more complicated then the procedure of perturbative quantization. Parallels can be drawn with the classical physics: linear equations (for example, Maxwell equations) are fundamentally simpler then nonlinear equations (Yang - Mills, Einstein equations). In the first case the general solutions of the corresponding equations are known whereas in the second case only some special solutions had been found (for example, non-Abelian monopoles, instantons and so on).

Here we would like to show that between nonperturbative quantization technique and turbulence modeling there is the analogy. 

\section{Initial equations for nonperturbative quantum field theory and turbulent: comparison}

Below we will see that nonperturbative quantization technique for quantum field theory (which is based either on an operator equation (that probably can not be solved) or on an infinite equations set for all Green functions) and modeling a turbulent flow with cumulants have much in common. It is well known that infinite equations set is used in turbulence theory for all cumulants (for the details one can see textbook \cite{Wilcox}).

\subsection{Equations}

In this section we present initial equations for nonperturbative quantum field theory (on the example of quantum chromodynamics) and for a flow of a turbulent fluid. For the reader convenience we split the text on two columns: on the left side the reader see everything concerning nonperturbative quantization and on the right side - turbulence.

\vspace{1cm}

\begin{minipage}[t]{0.5\textwidth}
\setlength{\rightskip}{10pt}
\begin{center}
\textbf{Quantum chromodynamics}
\end{center}

The classical SU(N) $(N = 1,2, \cdots , N)$ Yang-Mills equations are
  \begin{equation}
      \partial_\nu F^{B\mu\nu} = 0
  \label{tb1left-10}
  \end{equation}
  where $F^B_{\mu \nu} = \partial_\mu A^B_\nu -
  \partial_\nu A^B_\mu + g f^{BCD} A^C_\mu A^D_\nu$
  is the field strength; $B,C,D = 1, \ldots ,N$ are the SU(N) color indices; $g$ is the coupling constant; $f^{BCD}$ are the structure constants for the SU(N) gauge group. In quantizing the system given in equation \eqref{tb1left-10} - via Heisenberg's method \cite{heisenberg} one first replaces the classical fields by field operators $A^B_{\mu} \rightarrow \widehat{A}^B_\mu$. This yields the following differential equations for the operators
  \begin{equation}
    \partial_\nu \widehat {F}^{B\mu\nu} = 0.
  \label{tb1left-20}
  \end{equation}
  Equation \eqref{tb1left-20} can be considered as principal equation for the quantization of strongly interacting SU(3) gauge fields.
\end{minipage}
\begin{minipage}[t]{0.47\textwidth}
\begin{center}
\textbf{Turbulence}
\end{center}

In principle, the time-dependent, three-dimensional Navier-Stokes equation contains all of the physics of a given turbulent flow (on the right column we will follow to textbook \cite{Wilcox})
  \begin{equation}
    \rho \left(
     \frac{\partial v_i}{\partial t}
    + v_j \cdot \frac{\partial v_i}{\partial x_j}
    \right) = - \frac{\partial p}{\partial x_i} +
    \frac{\partial t_{ij}}{\partial x_j}
  \label{tb1right-30}
  \end{equation}
  where $v_i$ is the flow velocity, $\rho$ is the fluid density, $p$ is the pressure, $t_{ij} = 2 \mu s_{ij}$ is the viscous stress tensor,
  $s_{ij} = \frac{1}{2} \left(
    \frac{\partial v_i}{\partial x_j} + \frac{\partial v_j}{\partial x_i}
  \right)$, and $\mu$ is molecular viscosity.
\end{minipage}

\vspace{1cm}

The principal equations here are \eqref{tb1left-20} and \eqref{tb1right-30} that after averaging gives rise to an infinite equations set either for all Green's functions (for quantum field theory) or for all cumulants (for turbulence theory).

\subsection{How one can do the calculations}

In this section we would like to show that the calculations in nonperturbative quantum field theory and in turbulence theory are similar and have similar mathematical problems. We use $\langle \cdots \rangle$ for quantum averaging and $\overline{(\cdots)}$ for statistical one.

\vspace{1cm}

\begin{minipage}[t]{0.5\textwidth}
\setlength{\rightskip}{10pt}
\begin{center}
\textbf{Quantum chromodynmics}
\end{center}

The nonlinear operator equation \eqref{tb1left-20} for the field operators of the nonlinear quantum fields can be used to determine expectation values for the field operators
$\widehat {A}^B_\mu$, where
$\langle \cdots \rangle = \langle Q | \cdots | Q \rangle$ and $| Q \rangle$ is some quantum state). One can also use these equations to determine the expectation values of operators
that are built up from the fundamental operators $\widehat {A}^B_\mu$. The simple gauge field expectation values, $\langle A_\mu (x) \rangle$, are  obtained by average Eq. \eqref{tb1left-20} over some quantum state $| Q \rangle$
\begin{equation}
  \left\langle Q \left|
  \partial_\nu \widehat F^{B\mu\nu}
  \right| Q \right\rangle = 0.
\label{tb2left-10}
\end{equation}
One problem in using these equations to obtain expectation values like $\langle A^B_\mu \rangle$, is that these equations involve not only powers or derivatives of $\langle A^B_\mu \rangle$ ({\it i.e.} terms like $\partial_\alpha \langle
A^B_\mu \rangle$ or $\partial_\alpha
\partial_\beta \langle A^B_\mu \rangle$) but also contain terms like $\mathcal{G}^{BC}_{\mu\nu} = \langle A^B_\mu A^C_\nu \rangle$. Starting with Eq. \eqref{tb1left-20} one can generate an operator differential equation for the product $\widehat A^B_\mu \widehat A^C_\nu$ thus allowing the determination of the Green's function $\mathcal{G}^{BC}_{\mu\nu}$
\begin{equation}
  \left\langle Q \left|
  \widehat A^B(x) \partial_{y\nu} \widehat F^{B\mu\nu}(y)
  \right| Q \right\rangle = 0.
\label{tb2left-20}
\end{equation}
However this equation will in turn contain other, higher order Green's functions \eqref{tb2left-60}. Repeating these steps leads to an infinite set of equations connecting Green's functions of ever increasing order. In fact these equations are the Dyson-Schwinger equatons but ordinary the designation ``Dyson-Schwinger equatons '' is reserved for the application in perturbative quantum field theory. Finally we will have following equations set
\begin{eqnarray}
  \left\langle Q \left|
  \widehat A^B(x_1) \text{Eq. \eqref{tb1left-20}}
  \right| Q \right\rangle &=& 0 ,
\label{tb2left-30}\\
  \left\langle Q \left|
  \widehat A^B(x_1) A^B(x_2) \text{Eq. \eqref{tb1left-20}}
  \right| Q \right\rangle &=& 0 ,
\label{tb2left-40}\\
  \cdots
\nonumber \\
  \left\langle Q \left|
  \widehat A^B(x_1) \cdots A^B(x_n) \text{Eq. \eqref{tb1left-20}}
  \right| Q \right\rangle &=& 0 ,
\label{tb2left-50}
\end{eqnarray}
We consider these equations in nonperturbative quantum field theory. This construction, leading to an infinite set of coupled, differential equations, does not have an exact, analytical solution and so must be handled using some approximation.
\end{minipage}
\begin{minipage}[t]{0.47\textwidth}
\begin{center}
\textbf{Turbulence}
\end{center}

Because turbulence consists of random fluctuations of the various flow properties, a statistical approach for turbulence modeling is used. For a complete statistical description of the hydrodynamic fields of a turbulent flow it is required to have all the multidimensional joint probability distribution for the values ​​of these characteristics in space-time. But the definition of multivariate distributions is a very complex problem, in addition, these distributions are themselves often inconvenient for applications due to its awkwardness. Therefore, in practice one can form the time average of the continuity and Navier-Stokes equations. The nonlinearity of the Navier-Stokes equation leads to the appearance of momentum fluxes that act as apparent stresses throughout the flow. After that it is necessary to derive equations for these stresses and the resulting equations include additional unknown quantities. This illustrates the issue of closure, i.e., establishing a sufficient number of equations for all of the unknowns cumulants (moments).

\begin{center}
\textbf{Reynolds Averaging}
\end{center}

One can introduce the instantaneous velocity, $v_i(\vec x, t)$, as the sum of a mean, $V_i(\vec x, t)$, and a fluctuating part, $v^\prime_i(\vec x, t)$, so that
\begin{equation}
  v_i(\vec x, t) = V_i(\vec x, t) + v^\prime_i(\vec x, t).
\label{tb2right-10}
\end{equation}
The Reynolds averaged equations of motion is
\begin{equation}
  \rho \frac{\partial V_i}{\partial t} +
  \rho V_j \frac{\partial V_i}{\partial x_j}
  = - \frac{\partial p}{\partial x_i} +
  \frac{\partial }{\partial x_j} \left(
    2 \mu S_{ji} - \overline{\rho v_j^\prime v_i^\prime}
  \right)
\label{tb2right-20}
\end{equation}
Equation \eqref{tb2right-20} is usually referred to as the Reynolds-averaged Navier-Stokes equation. The quantity $\overline{\rho v_j^\prime v_i^\prime}$ is known as the Reynolds-stress tensor and denoted as
\begin{equation}
  \tau_{ij} = - \overline{\rho v_i^\prime v_j^\prime}.
\label{tb2right-30}
\end{equation}
Immediately we see that we have additionally six unknown quantities $\tau_{ij}$ as a result of Reynolds averaging. It is necessary to note that we have gained no additional equations. For general three-dimensional flows, we have four unknown mean-flow properties: pressure $p$ and the three velocity components $v_i$.
\end{minipage}

\begin{minipage}[t]{0.5\textwidth}
\setlength{\rightskip}{10pt}
One possible way to solve approximately equations \eqref{tb2left-30}-\eqref{tb2left-50} is following. We decompose n$-th$ Green function
\begin{equation}
	G_{\mu_1, \cdots , \mu_n}^{B_1, \cdots , B_n}(x_1, x_2 \cdots , x_n) =
	\left\langle
		A_{\mu_1}^{B_1}(x_1) \cdots A_{\mu_n}^{B_n}(x_n)
	\right\rangle
\label{tb2left-60}
\end{equation}

on the linear combination of the Green functions products of lower orders
\begin{equation}
\begin{split}
	& G_n(x_1, x_2 \cdots , x_n) \approx
\\
  &
  G_{n-2}(x_3, x_4 \cdots , x_n)
	\left[ G_2(x_1, x_2) -C_2 \right] +
\\
	&
	\left( \text{permutations of $x_1,x_2$ with
	$x_3, \cdots , x_n$} \right) +
\\
	&	
	G_{n-3} \left( G_3 -C_3 \right) + \cdots
\end{split}
\label{tb2left-70}
\end{equation}
where $C_{2,3 \cdots}$ are constants. In such a way one can cut off the infinite equation set \eqref{tb2left-30}-\eqref{tb2left-50}.

Another way to solve approximately the infinite equation set \eqref{tb2left-30}-\eqref{tb2left-50} is choose some functional (for instance, the action or something like to gluon condensates in quantum chromodynamics \footnote{in quantum chromodynamics there exist various gluon condensates:
$\mathrm{tr}\left(F_{\mu \nu} F^{\mu \nu} \right)$,
$\mathrm{tr} \left( F_\mu^\nu F_\nu^\rho F_\rho^\mu \right)$ and so on, for the review on the gluon condensate one can see Ref. \cite{Zakharov:1999jj}. Using these condensates and integrating their one can obtain various functionals.}) and write down its average expression using corresponding Green functions. After that it is necessary to use some well-reasoned physical assumptions to express the highest order Green function through Green function of lower order, see the decomposition \eqref{tb2left-70}. Finally we will have a functional which can be used to obtain Euler - Lagrange field equations for Green functions.
\end{minipage}
\begin{minipage}[t]{0.47\textwidth}
Along with the six Reynolds-stress components, we thus have ten unknown quantities. Our equations are mass conservation and the three components of equation \eqref{tb2right-20}. This means that our mathematical description of turbulent flow is not yet closed. To close the system, we must find enough equations to solve for our unknowns.

To obtain additional equations, we can take moments of the Navier-Stokes equation. That is, we multiply the Navier-Stokes equation by a fluctuating property and average the product. Using this procedure, we can derive, for example, a differential equation for the Reynolds-stress tensor.

One can obtain following equation for the Reynolds stress tensor
\begin{equation}
\begin{split}
  & \frac{\partial \tau_{ij}}{\partial t} +
  V_k \frac{\partial \tau_{ij}}{\partial x_k} =
  - \tau_{ik} \frac{\partial V_j}{\partial x_k} -
  \tau_{jk} \frac{\partial V_i}{\partial x_k} +
\\
  &
  2 \overline{\mu \frac{\partial v^\prime_i}{\partial x_k}
  \frac{\partial v^\prime_j}{\partial x_k}} +
  \overline{v^\prime_i \frac{\partial p^\prime}{\partial x_j}} +
  \overline{v^\prime_j \frac{\partial p^\prime}{\partial x_i}} +
\\
  &
  \frac{\partial}{\partial x_k} \left(
    \nu \frac{\partial \tau_{ij}}{\partial x_k} +
    \overline{\rho v^\prime_i \rho v^\prime_j \rho v^\prime_k}
  \right)
\label{tb2right-40}
\end{split}
\end{equation}
We have obtained six new equations, one for each independent component of the Reynolds-stress tensor $\tau_{ij}$. However, we have also generated 22
new unknowns $\overline{\rho v^\prime_i \rho v^\prime_j \rho v^\prime_k}$, $\overline{\mu \frac{\partial v^\prime_i}{\partial x_k}
\frac{\partial v^\prime_j}{\partial x_k}}$ and
$\overline{v^\prime_i \frac{\partial p^\prime}{\partial x_j}}$,
$\overline{v^\prime_j \frac{\partial p^\prime}{\partial x_i}}$.

It illustrates the closure problem of turbulence. Because of the nonlinearity of the Navier-Stokes equation, as we take higher and higher moments, we generate additional unknowns at each level. The function of turbulence modeling is to devise approximations for the unknown correlations in terms of flow properties that are known so that a sufficient number of equations exists. In making such approximations, we close the system.

There exist a few approaches to close this infinite equations set. One of them are: algebraic (zero-equation) models, one-equation models, two-equation models and second-order closure models.
\end{minipage}

\vspace{1cm}

Comparing left and right columns we see that the procedure of obtaining corresponding infinite equations set for both cases is practically the same with the difference concerning physical quantities: quantum fields for the left side and velocities, pressure and so on for the right side.

\section{Interweaving of Planck constant, Reynolds number and coupling constant}

Above we have seen that the procedures of nonperturbative quantization and turbulence modeling have common properties. In this section we will show that there exists an interweaving of Planck constant, Reynolds number and coupling constant (here we follow to Ref. \cite{Dzhunushaliev:2009turb-qcd}).

In Ref. \cite{Dzhunushaliev:2009turb-qcd} it is found interesting relation between coupling constant in quantum field theory and Reynolds number in hydrodynamics.

\vspace{1cm}

\begin{minipage}[t]{0.5\textwidth}
\setlength{\rightskip}{10pt}
In quantum field theory there are perturbative regime when a dimensionless coupling constant
\begin{equation}
	\alpha^2 = \frac{1/\tilde g^2}{\hbar c}
\label{3-20}
\end{equation}
is small enough $\alpha^2 < 1$ and nonperturbative regime when $\alpha^2 \geq 1$ (here $\tilde g$ is a dimension coupling constant, $\hbar$ is Planck constant and $c$ is the speed of light). In quantum field theory the open question is a nonperturbative quantization for $\alpha^2 \geq 1$. For example, the fine-structure constant in quantum electrodynamics is
$\alpha^2 = e^2/\hbar c \approx 1/137$ ($e$ is the electron charge), $\alpha^2 \approx 0.1$ for a weak interaction and $\alpha^2 \geq 1$ for a strong interaction.
\end{minipage}
\begin{minipage}[t]{0.47\textwidth}
As we know the character of fluid flow depends on Reynolds number $\mathrm{Re}$
\begin{equation}
	\mathrm{Re} = \frac{\rho v l}{\mu}
\label{3-10}
\end{equation}
where $l$ is a characteristic length for a given flow. If $\mathrm{Re} < \mathrm{Re}_{cr}$ then the flow is laminar one, if $\mathrm{Re} > \mathrm{Re}_{cr}$ then the flow is turbulent one.
\end{minipage}

\vspace{1cm}

If we rewrite equation \eqref{3-10} as
\begin{equation}
	\mathrm{Re} = \frac{\rho v^2 l^4}{\mu l^3 v}
\label{3-30}
\end{equation}
then we have following dimensional equalities:
$\left[ \rho v^2 l^4 \right] = \left[ 1/\tilde g^2 \right] = g \cdot cm^3 / s^2$,
$\left[ \hbar \right] = \left[ \mu l^3 \right] = g \cdot cm^3 / s$. It allow us to offer following relations
\begin{eqnarray}
	1/\tilde g^2 & \leftrightarrow & \rho v^2 l^4 ,
\label{3-40}\\
	\hbar & \leftrightarrow & \mu l^3 .
\label{3-50}
\end{eqnarray}
That allows us to do following analogies:
\begin{itemize}
  \item $\alpha^2 \leftrightarrow \mathrm{Re}$.
	\item $\hbar = 0 \leftrightarrow \mu = 0$. In this case a classical theory corresponds to an ideal fluid.
	\item $\hbar \neq 0 \text{ and } \alpha^2 < 1 \leftrightarrow \mu \neq 0 \text{ and } \mathrm{Re} < \mathrm{Re}_{cr}$. In this case a laminar fluid corresponds to a perturbative regime of quantum field theory.
	\item $\hbar \neq 0 \text{ and } \alpha^2 \geq 1 \leftrightarrow \mu \neq 0 \text{ and } \mathrm{Re} > \mathrm{Re}_{cr}$. In this case a turbulent fluid corresponds to a nonperturbative regime of quantum field theory.
\end{itemize}

\section{Discussion}

Thus we have shown that the main similarity between nonperturbative quantization technique and turbulence modeling is an infinite equations set (either \eqref{tb2left-30} - \eqref{tb2left-50} for the first case or \eqref{tb2right-20} \eqref{tb2right-40} for the second case.) The main problem in solving these equations set is a closure problem: cutting off of the infinite equations set. For such truncation it is necessary to bring into account some physical assumptions. For example, two scalar field approximation for quantum chromodynamics \cite{Dzhunushaliev:2011we} or algebraic (zero-equation) models, one-equation models, two-equation models and second-order closure models for turbulence modeling.

One principal difference between nonperturbative quantization and turbulence is that equations set \eqref{tb2left-30} - \eqref{tb2left-50} is equivalent to one operator equation \eqref{tb1left-20} but it is not true for the turbulence.

It is necessary to note that this truncation may lead to physically incorrect results (negative values ​​of positive definite quantities: probability density, dispersions, energy dissipation and so on). The latter is due to the fact that if we specify a finite number of cumulants (moments) then the probability distribution may not exist. It is closely related to the inadmissibility of arbitrary truncation of the Taylor series of the characteristic function.

One can also note that in some cases quantum chromodynamics behaves like a normal or ideal fluid \cite{Torrieri:2011ne} \cite{Endlich:2010hf}.

\section*{Acknowledgements}

This work was partially supported  by a grant in fundamental research in natural sciences by Science Committee of the Ministry of Education and Science of Kazakhstan and by a grant of VolkswagenStiftung.


\begin{thebibliography}{99}

\bibitem{heisenberg}
W. Heisenberg, \textit{Introduction to the unified field theory of
elementary particles.}, Max - Planck - Institut f\"ur Physik und
Astrophysik, Interscience Publishers London, New York, Sydney,
1966.

\bibitem{Wilcox}
David С Wilcox, \emph{Turbulence Modeling for CFD}, DCW Industries, Inc.
La Canada, California, 1994.

\bibitem{Dzhunushaliev:2009turb-qcd}
  V.~Dzhunushaliev,
	``Flux tube in turbulent flow and quantum chromodynamics,''
  AMCOS, Vol. 1, Issue 1, 2010.
	arXiv:0907.3624 [physics.flu-dyn].


\bibitem{Dzhunushaliev:2011we}
  V.~Dzhunushaliev,
  ``SU(3) glueball gluon condensate,''
  [arXiv:1110.1427 [hep-ph]];\\
  V.~Dzhunushaliev,
  ``SU(3) flux tube gluon condensate,''
  arXiv:1010.1621 [hep-ph].

\bibitem{Zakharov:1999jj}
  V.~I.~Zakharov,
  Int.\ J.\ Mod.\ Phys.\ A {\bf 14}, 4865 (1999)
  [hep-ph/9906264].

\bibitem{Torrieri:2011ne} 
  G.~Torrieri,
  Phys.\ Rev.\ D {\bf 85}, 065006 (2012)
  [arXiv:1112.4086 [hep-th]].

\bibitem{Endlich:2010hf} 
  S.~Endlich, A.~Nicolis, R.~Rattazzi and J.~Wang,
  JHEP {\bf 1104}, 102 (2011)
  [arXiv:1011.6396 [hep-th]].
  


\end{thebibliography}
\end{document}